\documentclass[prd,showpacs,showkeys,amsmath,amssymb]{revtex4}

\usepackage{graphicx}

\newcommand{\bfbeta}{\mbox{\boldmath $\beta$}}
\newcommand{\bftheta}{\mbox{\boldmath $\theta$}}
\newcommand{\bfu}{\mbox{\boldmath $u$}}
\newcommand{\bfalpha}{\mbox{\boldmath $\alpha$}}

\def\ls{\mathrel{\lower0.6ex\hbox{$\buildrel {\textstyle <}\over{\scriptstyle \sim}$}}}

\begin{document}

\author{V. Bozza$^{a,b}$, M. Sereno$^{c}$}

\affiliation{
$^a$ Dipartimento di Fisica `E.R. Caianiello', Universit\`{a} di Salerno, via Allende, I-84081 Baronissi (SA), Italy.
\\
$^b$  Istituto Nazionale di Fisica Nucleare, Sezione di Napoli.
\\
$^c$ Institut f\"{u}r Theoretische Physik der Universit\"{a}t Z\"{u}rich, Winterthurerstrasse 190, CH-8057 Z\"{u}rich, Switzerland
}

\title{The weakly perturbed Schwarzschild lens in the strong deflection limit}

\date{10 January 2006}

\begin{abstract}
We investigate the strong deflection limit of gravitational
lensing by a Schwarzschild black hole embedded in an external
gravitational field. The study of this model, analogous to the Chang
\& Refsdal lens in the weak deflection limit, is important to
evaluate the gravitational perturbations on the relativistic
images that appear in proximity of supermassive black holes hosted
in galactic centers. By a simple dimensional argument, we prove
that the tidal effect on the light ray propagation mainly occurs
in the weak field region far away from the black hole and that the
external perturbation can be treated as a weak field quadrupole
term. We provide a description of relativistic critical curves and
caustics and discuss the inversion of the lens mapping.
Relativistic caustics are shifted and acquire a finite diamond
shape. Sources inside the caustics produce four sequences of
relativistic images. On the other hand, retro-lensing caustics are
only shifted while remaining point-like to the lowest order.
\end{abstract}



\maketitle

\section{Introduction}

Gravitational lensing in the Strong Deflection Limit (SDL) is
emerging as a promising tool for black hole investigation. It is
well known that photons making one or more complete loops around
the black hole before approaching the observer produce two
infinite series of images very close to the shadow of the black
hole \cite{dar59,cha83,oha89}. However, this fact became an
astrophysical breakthrough only when the super-massive black hole
supposed to be hosted in the radio-source Sgr~A* in the Galactic
center was proposed as an ideal candidate for detecting
relativistic images of sources passing behind it \cite{vi+el00}.
SDL investigations strongly benefitted from expansion techniques
of the deflection angle \cite{boz+al01}, which were then
generalized to arbitrary spherically symmetric metrics
\cite{boz02}. Theoretical properties of SDL lensing are now well
assessed for a large class of black holes, since effects of charge
\cite{eir+al02,zac+al05}, rotation \cite{boz03,boz+al05} and
several extended theories of gravitation
\cite{boz02,bha03,eir05,whi05,ma+mu05,te+la05,eir05b,na+al06,sa+bh06}
have been considered.

On the observational side, lensing effects are expected to play a
significant role in all high-resolution observations. However, a
clean detection of a deeply relativistic image of some source is
by no way an easy task, since Sgr~A* must be resolved at least at
microarcsecond level, and a source is needed with a surface
brightness much higher than Sgr~A* itself in the spectral band in
which observations are lead. The development of Very Long Baseline
Interferometry (VLBI) techniques in different bands of the
spectrum is fast approaching the microarcsecond resolution
required to distinguish such relativistic images (for an updated
discussion on observational perspectives of relativistic images,
see e.g. Ref. \cite{bo+ma05}). As discussed in Ref.
\cite{boz+al05}, the best prospect comes from relativistic images
of low mass X-ray binaries around Sgr~A* \cite{mun+al05}, which
should be detected by future X-ray interferometry missions such as
MAXIM (http://maxim.gsfc.nasa.gov).

Supermassive black holes should be hosted in galactic bulges and
the surrounding environment could perturb the space-time of the
main lens. As the black hole is part of a larger system, the
gravitational field around it can not be exactly spherically
symmetric. In some cases, large clusters of stars or secondary
black holes are known to lie close to supermassive black holes. In
this paper, we want to discuss the effect of external
perturbations on the trajectories of light rays that are highly
deflected. In the weak deflection limit, lenses with perturbed
symmetry are usually modelled as quadrupole lenses \cite{sef}. If
the gravitational field of the perturber changes very little over
the relevant length scale of the black hole, it is enough to
expand the potential of the external macro-lens up to the first
non trivial term, i.e. the quadratic term. The perturbed
Schwarzschild lens in the weak deflection limit is known as
Chang-Refsdal lens after its first investigators
\cite{ch+re79,ch+re84}. It was also recently revisited in Ref.
\cite{an+ev06}. Here, we offer a complementary treatment in the
SDL, which may be relevant to clarify the role of external fields
in the formation of relativistic images. The paper is structured
as follows. In Section~\ref{sec:lenseq}, we introduce the general
lens equation for an unperturbed Schwarzschild black hole and
specify it for the two cases of standard lensing and
retro-lensing. In Section~\ref{sec:defl} we analyze the
perturbation on the deflection angle exerted by an external weak
field, showing that it mainly occurs in the weak field regime. In
Section~\ref{sec:standard}, we analyze the perturbed lens equation
in the standard lensing geometry and in the retro-lensing
configuration, finding critical curves, caustics and images.
Section~\ref{sec:conc} is devoted to some final considerations.

\section{The Schwarzschild lens in the strong deflection limit}
\label{sec:lenseq}

If the lens is spherically symmetric, as in the case of a
Schwarzschild black hole, the motion of the photon takes place on
a single plane. Then the lens equation relating the angular
position of the source $\beta$ and the angular position of a
lensed image $\theta$, relative to the position of the lens in the
sky, can be found by simple trigonometry, looking at Fig. \ref{Fig
lens}.

\begin{figure}
\resizebox{\hsize}{!}{\includegraphics{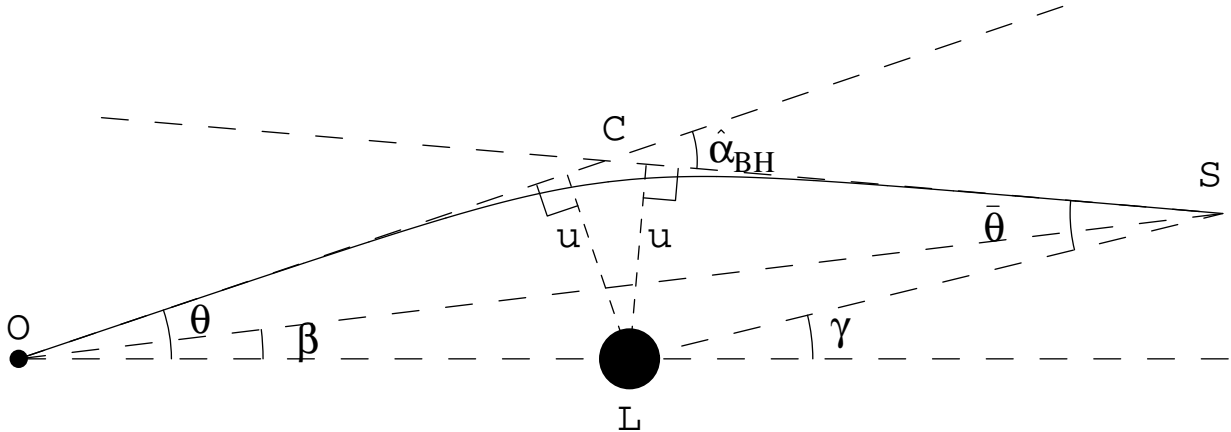}} \caption{Generic
lensing configuration.}
 \label{Fig lens}
\end{figure}

We denote by $D_\mathrm{L}$, $D_\mathrm{S}$ and $D_\mathrm{LS}$
the distance between observer and lens, observer and source, and
lens and source respectively. They are related by Carnot theorem
$D_\mathrm{LS}^2=D_\mathrm{L}^2+D_\mathrm{S}^2-2
D_\mathrm{L}D_\mathrm{S}\cos\beta$. We also introduce the angle
$\gamma$ as the angular position of the source as seen from the
lens. The angle $\overline{\theta}$ is the emission angle of the
light ray from the source w.r.t the lens. The deflection angle
$\hat\alpha_\mathrm{BH}$ is the angle between the asymptotic
incident and outgoing directions of the photon.

The impact parameter $u$ can be taken as the distance between the
lens and the incident direction or the outgoing direction. The two
distances are indeed the same by time-reversal symmetry. This
implies the equality
\begin{equation}
D_\mathrm{L}\sin \theta =D_\mathrm{LS} \sin \overline{\theta}.
\label{thetabar}
\end{equation}
Let us consider the triangle $OLS$. Indicating the angle
$\widehat{OSL}$ by $\zeta$, the sum of the internal angles gives
\begin{equation}
\zeta +\beta+(\pi-\gamma)=\pi.
\end{equation}
Considering the triangle $OCS$, the sum of the internal angles
gives
\begin{equation}
(\theta-\beta) + (\overline{\theta}-\zeta) +(\pi-\hat\alpha_\mathrm{BH})= \pi.
\end{equation}
Summing the two equations, we get the relation
\begin{equation}
\gamma=\theta+\overline{\theta}-\hat\alpha_\mathrm{BH},
\label{gamma}
\end{equation}
which has already been introduced as lens equation in Refs.
\cite{boz03,bo+ma04,boz+al05}. Substituting this expression for
$\gamma$ in the equality $D_\mathrm{S} \sin \beta =D_\mathrm{LS}
\sin \gamma$, we get
\begin{equation}
D_\mathrm{S} \sin \beta = D_\mathrm{LS} \sin \overline{\theta}
\cos(\hat\alpha_\mathrm{BH} -\theta)-D_\mathrm{LS}\cos
\overline{\theta} \sin(\hat\alpha_\mathrm{BH} -\theta).
\end{equation}
Using Eq. (\ref{thetabar}) to eliminate $\overline{\theta}$, we
finally get the general form of the lens equation
\begin{equation}
D_\mathrm{S} \sin \beta = D_\mathrm{L} \sin \theta
\cos(\hat\alpha_\mathrm{BH}
-\theta)-\sqrt{D_\mathrm{LS}^2-D_\mathrm{L}^2\sin^2\theta}
\sin(\hat\alpha_\mathrm{BH} -\theta). \label{cr1}
\end{equation}

This equation is more general than the lens equation derived in
Ref.~\cite{vi+el00}, which was derived assuming that $C$ belongs
to the lens plane. That assumption is only valid for small
$\hat\alpha_\mathrm{BH}$, making that equation inadequate for
retro-lensing or other cases when the incident direction differs
very much from the outgoing direction. Our lens equation is
completely general and can be used to describe any configuration.

If the lens is an unperturbed Schwarzschild black hole of mass $m$, the
deflection angle takes the classical weak field form $4m/u$ for large impact parameters $u$, where we are using geometrized units (gravitational constant $G=1$, speed of light in vacuum $c=1$). For small impact parameters, the
deflection angle grows more and more, diverging at
$u=u_\mathrm{m}=3\sqrt{3}m$. The calculation of the exact
deflection angle as a function of $u$ has been first carried out by \citet{dar59} analytically in terms of elliptic integrals. A
good approximation, valid for small impact parameters (large
deflection angles), is the so-called Strong Deflection Limit
(SDL), firstly introduced by Darwin himself and lately generalized
by \citet{boz02} to any spherically symmetric black hole . It
reads
\begin{equation}
\label{cr4} \hat{\alpha}_\mathrm{BH} \simeq -\bar{a} \ln\left(
\frac{u}{u_\mathrm{m}} -1 \right) + \bar{b} ,
\end{equation}
For a Schwarzschild black hole, $\bar{a} = 1$, $\bar{b}
=-0.40002$. Note that $\theta_\mathrm{m}\simeq
u_\mathrm{m}/D_\mathrm{L}$ corresponds to the angular radius of
the apparent shadow of the black hole, which is typically very
small in all cases of interest. For example, it is expected to be
$ \sim 23~\mu$as for the supermassive black hole at the center of our
Galaxy. This formula is strictly valid for
$D_\mathrm{LS},D_\mathrm{L} \gg 2m$. For sources or observers very
close to the black hole it needs to be corrected.

As shown in several papers
\cite{dar59,oha89,nem93,vi+el00,boz+al01,boz02,ei+to04},
relativistic images formed by light rays experiencing large
deflections are prominent only when the source is in proximity of
a caustic point. The caustic points are behind the lens and in
front of the lens. The first class of caustic points is relevant
for standard gravitational lensing configurations, in which the
source is almost aligned behind the lens. The second class is
relevant for retro-lensing configurations, where the source is on
the same side of the observer and the light rays must turn around
the black hole in order to hit the observer. In the following, we
shall specify the general lens equation for sources close to these
two classes of caustic points.

\subsection{Lens equation for standard lensing geometry}

Considering sources very close to the optical axis behind the
black hole, we have $\beta,\gamma \ll 1$, while the relation
$D_\mathrm{L}\gg 2m$ implies $\theta, \overline{\theta} \ll 1$.
Then, Eq. (\ref{gamma}), interpreted as an equation valid modulo
$2\pi$, implies $\hat{\alpha}_\mathrm{BH} = 2n \pi + \Delta
\hat{\alpha}_n^\mathrm{BH}$, with $\Delta
\hat{\alpha}_n^\mathrm{BH} \ll 1$. In consequence of this, all
trigonometric functions in Eq. (\ref{cr1}) can be expanded for
small angles. Taking the first order terms of these expansions, we
get
\begin{equation}
\label{cr3}
\beta \simeq \theta - \frac{D_\mathrm{LS}}{D_\mathrm{S}}
 \Delta \hat{\alpha}_n^\mathrm{BH},
\end{equation}
where we have also used that $D_\mathrm{S}\simeq
D_\mathrm{L}+D_\mathrm{LS}$. This is the lens equation for standard lensing configurations. The
weak field lens equation is included for $n=0$, i.e. when the
deflection angle is very close to zero. Then $\Delta
\hat{\alpha}_{n=0}^\mathrm{BH}$ coincides with $
\hat{\alpha}_\mathrm{BH}$. On the other hand, for small impact
parameters, the deflection angle grows very large, making the
photon complete one or more loops around the black hole. In such
cases, in the lens equation one has to consider the offset $\Delta
\hat{\alpha}_n^\mathrm{BH}= \hat{\alpha}_\mathrm{BH}-2n\pi$.

If the source is behind the lens, we have images for each
integer $n$, corresponding to a deflection angle that is close to
a multiple of $2\pi$. In consequence of this, we are mainly
interested in those values of $\theta$ such that
$\hat\alpha_\mathrm{BH}$ is a multiple of $2\pi$. Indicating these values by $\theta_n^0$, we have
$\hat\alpha_{BH}(D_\mathrm{L}\theta_n^0)=2n\pi$. From Eq. (\ref{cr4}), we get a simple expression for $\theta_n^0$
\begin{equation}
\label{cr5}
\theta_n^0 \simeq \theta_\mathrm{m} (1+e_n), \ \
e_n \equiv e^{(\bar{b} -2n \pi)/\bar{a}}.
\end{equation}
Then, letting $\epsilon \equiv (\theta - \theta_n^0)/\theta_n^0$ and
expanding the off-set around $\theta_n^0$, we get
\begin{equation}
\label{cr6} \Delta \hat{\alpha}_n^\mathrm{BH} =
-\frac{\bar{a}}{e_n}\frac{\theta_n^0 }{\theta_\mathrm{m}}\epsilon
.
\end{equation}
The lens equation reduces to
\begin{equation}
\label{cr7} \beta \simeq \theta_n^0 \left( 1 -
\frac{\epsilon}{\epsilon_n^\mathrm{E}} \right),
\end{equation}
where
\begin{equation}
\label{cr8}
\epsilon_n^\mathrm{E} \equiv
-\frac{D_\mathrm{S}}{D_\mathrm{LS}}\frac{e_n}{\bar{a}}\theta_\mathrm{m}
.
\end{equation}
and we have already taken into account that $\epsilon,
\epsilon_n^\mathrm{E} \ll 1$ to get rid of terms of order
$\epsilon$.

Images on the same side of the source solve Eq.~(\ref{cr8}) for a
positive $\beta$. The solution of the same equation with the
source placed in $-\beta$ gives the position of the images on the
opposite side. If $\beta =0$, the source is lensed in an infinite
sequence of relativistic Einstein rings at $\theta =
\theta_n^0(1+\epsilon_n^\mathrm{E})$. For $\beta \neq 0$, two
infinite sequences of relativistic images form on opposite sides
of the black hole at
\begin{equation}
\label{cr9} \epsilon  =  \left( 1 \mp \frac{\beta}{\theta_n^0}
\right) \epsilon_n^\mathrm{E}.
\end{equation}
Note that for any relativistic image $\epsilon \approx
\epsilon_n^\mathrm{E} \approx
\frac{D_\mathrm{S}}{D_\mathrm{LS}}\theta_\mathrm{m}$. Thus, unless
$D_\mathrm{LS} \approx 2m$, we have $\epsilon \ll 1$, so that
$\theta_n^0$ already represents an excellent approximation to the
position of the $n$-th relativistic image.

\subsection{Lens equation for retro-lensing geometry}

The treatment presented up to now only covers standard lensing
geometry, with a source almost aligned behind the black hole. The
modifications needed to account for the case in which the source
is in front of the lens are very simple. In fact, redefining
$\theta_n^0$ by the equation
$\hat{\alpha}_\mathrm{BH}(D_\mathrm{L}\theta_n^0) = (2n-1) \pi$,
we get
\begin{equation}
\theta_n^0 = \theta_\mathrm{m} (1+e_n), \ \ e_n = e^{[\bar{b}
-(2n-1) \pi]/\bar{a}}.
\end{equation}
The expansion of Eq.~(\ref{cr1}) for small angles and
$\hat{\alpha}_\mathrm{BH} = (2n-1) \pi + \Delta
\hat{\alpha}_n^\mathrm{BH}$  leads to
\begin{equation}
\label{cr3re}
\beta \simeq -\left( 1+2
\frac{D_\mathrm{LS}}{D_\mathrm{S}} \right) \theta +
\frac{D_\mathrm{LS}}{D_\mathrm{S}}
 \Delta \hat{\alpha}_n^\mathrm{BH},
\end{equation}
which differs from Eq.~(\ref{cr3}) for two signs coming from the
expansions of the trigonometric functions and from the fact that
now $D_\mathrm{S}=D_\mathrm{L}-D_\mathrm{LS}$. Since the
magnification privileges sources close to the black hole, here we
only consider $D_\mathrm{LS}<D_\mathrm{L}$, while Ref.~\cite{ei+to04} considers the other case
$D_\mathrm{LS}>D_\mathrm{L}$, with a slightly different expression
for the lens equation.

The expression for $\Delta \hat{\alpha}_n^\mathrm{BH}$ remains the
same given in Eq. (\ref{cr6}), so that the lens equation can be
written as
\begin{equation}
\beta \simeq -\left( 1+2 \frac{D_\mathrm{LS}}{D_\mathrm{S}} \right)
\theta_n^0 \left(1- \frac{\epsilon}{\epsilon_n^E} \right),
\end{equation}
where now
\begin{equation}
\label{cr8re} \epsilon_n^\mathrm{E} \equiv
-\frac{D_\mathrm{S}}{D_\mathrm{LS}}\frac{e_n}{\bar{a}}\theta_\mathrm{m}
\left( 1+2 \frac{D_\mathrm{LS}}{D_\mathrm{S}} \right)^{-1}.
\end{equation}
The relativistic Einstein rings form at $\epsilon=\epsilon_n^E$.
For $\beta \neq 0$, the relativistic images form at
\begin{equation}
\label{cr9re} \epsilon =  \left[ 1 \pm
\frac{\beta}{\theta_n^0}\left( 1+2
\frac{D_\mathrm{LS}}{D_\mathrm{S}} \right)^{-1} \right]
\epsilon_n^\mathrm{E}.
\end{equation}

\section{Deflection angle in a perturbed Schwarzschild geometry}
\label{sec:defl}

We assume that the perturbation is weak, so that the gravitational
potential  $\varphi_\mathrm{Q}$ induced by the tidal field is
small throughout the photon trajectory. The order of the curvature
radius $\mathcal{R}$ of the external field at the black hole
position is then determined by the scale of the variations in the
gravitational potential induced by the perturber. Indicating this
scale by $b$, we can thus establish the relation
\begin{equation}
\mathcal{R}^{-2} \approx \partial_i
\partial_j  \varphi_\mathrm{Q} \approx b^{-2} | \varphi_\mathrm{Q}|.
\label{R-2}
\end{equation}
If the scale of the variations $b$ is much larger than the
Schwarzschild radius of the black hole $2m$, this relation
automatically implies $\mathcal{R}\gg 2m$. For example, if the
external field is due to a second point-mass, we have
$\varphi_\mathrm{Q}=-2m_2/r_2$, where $r_2$ is the distance of the
generic point from the second mass. In this case, the scale of the
variation of such a potential around the first black hole is
dictated by the distance $b$ between the two black holes and
$\mathcal{R}^{-2}\approx m_2/b^3$, coherently with what we know
from the curvature invariants of the Schwarzschild metric.

Coming back to the general situation, the metric of a spherically
symmetric black hole distorted by an external gravitational field
can be calculated with an expansion in inverse powers of the local
radius of curvature $\mathcal{R}$ of the external spacetime. The
first corrections, proportional to $\mathcal{R}^{-2}$, have been
calculated by \citet{alv00}, who found an approximate form of the
metric of a binary system of black holes. In particular, he
divided the spacetime in four regions: two of them surround the
two black holes, an intermediate region is then encircled by the
asymptotic region far from the two black holes. For each region,
he found an approximated expression for the metric, nicely
matching the metrics of the neighbor regions in the overlap zones.
These approximate solutions were proposed as initial conditions
for numerical programs designed to find the best approximation to
the whole spacetime. Recently, \citet{poi05}  gave the general
expression for the tidal distortion of a Schwarzschild black hole
embedded in an arbitrary spacetime up to terms proportional to
$\mathcal{R}^{-3}$. In his work, the external spacetime determines
the tidal fields through the electric and magnetic components of
the Weyl tensor. These fields are then multiplied by radial
functions to be determined solving the Einstein equations. The
metric found by Poisson reproduces Alvi's metric when the tidal
fields are generated by a second Schwarzschild black hole. What
matters for us, as we shall see in a while, is that the first
corrections close to the black hole are always of order of
$\mathcal{R}^{-2}$.

In an asymptotically flat spacetime the deflection vector is
simply given by the difference between the incoming and outgoing
tangent vectors to the photon path. Integrating the geodesics equation, we get
\begin{equation}
\hat\bfalpha=\dot x^i_\mathrm{in}-\dot x^i_\mathrm{out}=\int
{\Gamma^i}_{\mu\nu} \dot x^\mu \dot x^\nu d\tau, \label{DeflVec}
\end{equation}
where ${\Gamma^i}_{\mu\nu}$ is the affine connection and $\tau$ is an affine parameter. The
integral can be split into three pieces: one covering the approach
to the black hole, where the effects due to the black hole and the
perturber are both weak; a second piece where the photon winds
around the tidally distorted black hole under its strong
gravitational field; a third piece where the photon goes away from
the black hole and the deflections due to black hole and perturber
are again weak. Denoting by $-\tau_\mathrm{Strong}$ and $\tau_\mathrm{Strong}$ the values of
the affine parameter where the transitions between the different
regimes take place, we have
\begin{eqnarray}
\hat\bfalpha & =& \hat\bfalpha_\mathrm{in}+\hat\bfalpha_\mathrm{Strong}+\hat\bfalpha_\mathrm{out} ,
\\
\hat\bfalpha_\mathrm{in} & = & \int\limits^{-\tau_\mathrm{Strong}}_{-\infty}
{\Gamma^i}_{\mu\nu} \dot x^\mu \dot x^\nu d\tau , \\
\hat\bfalpha_\mathrm{Strong} & = & \int\limits^{\tau_\mathrm{Strong}}_{-\tau_\mathrm{Strong}}
{\Gamma^i}_{\mu\nu} \dot x^\mu \dot x^\nu d\tau , \\
\hat\bfalpha_\mathrm{out} & = & \int\limits_{\tau_\mathrm{Strong}}^{\infty}
{\Gamma^i}_{\mu\nu} \dot x^\mu \dot x^\nu d\tau.
\end{eqnarray}
The first and the third piece can be calculated in the weak field
regime, where ${\Gamma^i}_{\mu\nu} \dot x^\mu \dot x^\nu \simeq 2
( \nabla_\perp \varphi)^{i}$, with $\nabla_\perp \varphi$ being
the projection of the gradient onto the plane orthogonal to the
direction of the light ray and $\varphi$ the Newtonian potential,
given by the sum of the black hole and perturber contributions
$\varphi=\varphi_\mathrm{BH}+\varphi_\mathrm{Q}$. The central
piece, $\hat\bfalpha_\mathrm{Strong}$, can be expanded in inverse
powers of the tidal radius $\mathcal{R}$. Following Poisson
\cite{poi05}, the first order correction
$\hat\bfalpha_\mathrm{Strong}^{(1)}$ is of order of
$\mathcal{R}^{-2}$. We can thus write
\begin{eqnarray}
\hat\bfalpha & \simeq & 2\int\limits^{-\tau_\mathrm{Strong}}_{-\infty} \nabla_\perp
\varphi_\mathrm{BH} d\tau+ \hat\bfalpha_\mathrm{Strong}^{(0)}
+2\int\limits_{\tau_\mathrm{Strong}}^{\infty} \nabla_\perp \varphi_\mathrm{BH}
d\tau \nonumber \\
& +& 2\int\limits^{-\tau_\mathrm{Strong}}_{-\infty}
\nabla_\perp \varphi_\mathrm{Q} d\tau +
\hat\bfalpha_\mathrm{Strong}^{(1)} + 2\int\limits_{\tau_\mathrm{Strong}}^{\infty}
\nabla_\perp \varphi_\mathrm{Q} d\tau.
\end{eqnarray}
The first three pieces reconstruct the unperturbed Schwarzschild
deflection vector $\hat\bfalpha_\mathrm{BH}$, related to the
scalar deflection angle introduced in the previous section by
\begin{equation}
|\hat\bfalpha_\mathrm{BH}|= 2\left| \sin \frac{\hat
\alpha_\mathrm{BH}-2n\pi}{2} \right| .
\end{equation}
Note that the deflection vector, as defined by Eq.~(\ref{DeflVec}), is independent of the number of loops performed
by the photon around the black hole, since it only refers to the
asymptotic directions.

Now let us consider the terms introduced by the perturber. The perturbation in the metric of a tidally distorted black hole is of the order of $\mathcal{R}^{-2}$, i.e. $\delta g_{\mu \nu} \approx (r/\mathcal{R})^2 $ \cite{poi05}. Then, $\delta \Gamma^i _{\mu \nu}  \approx   (r/\mathcal{R}^2) $. The contribution to the deflection angle due to the external perturbation is then given by an integral over the strong field region, whose size is governed by the Schwarzschild radius of the black hole $2m$, of the perturbation to the affine connection in the tidally distorted Schwarzschild metric,
\begin{equation}
\hat\bfalpha_\mathrm{Strong}^{(1)}  \approx
\int\limits_{-2m}^{2m}  \left| \frac{\tau}{\mathcal{R}^2} \right|
d \tau
  \approx  \left( \frac{2 m}{\mathcal{R}} \right)^2
\end{equation}
This is what expected on a dimensional analysis since $\hat\bfalpha_\mathrm{Strong}^{(1)} \propto {\mathcal{R}}^{-2}$ and, apart from the local curvature radius of the external field providing the scale relevant for tidal perturbations, the only other length is $2 m$ itself.

On the other hand, the two weak field contributions contain
integrals of $\nabla_\perp \varphi_\mathrm{Q} \approx b^{-1}
\varphi_\mathrm{Q}$ and cover the whole photon trajectory, safe
for a small piece whose length is of order $2m$. Since the scale
of variations $b$ determines the size of the region where
$\varphi_\mathrm{Q}$ is relevant, the two weak field contributions
are of order
\begin{equation}
|\hat\bfalpha_\mathrm{Q}|=\left|
2\int\limits^{-\tau_\mathrm{Strong}}_{-\infty}\nabla_\perp \varphi_\mathrm{Q}
d\tau + 2\int\limits_{\tau_\mathrm{Strong}}^{\infty}\nabla_\perp
\varphi_\mathrm{Q} d\tau \right| \approx | \varphi_\mathrm{Q} |.
\end{equation}
Recalling Eq.~(\ref{R-2}), we see that the contribution to the deflection due to the perturber from the weak field region is much larger than the strong field
contribution.

We can thus conclude that the perturbation of the trajectory of a
photon passing close to a black hole due to an external field is
dominated by the weak field contribution and that the deviations
arising in the strong regime can be safely neglected. If we
replace the two integrals in the range
$]-\infty,-\tau_\mathrm{Strong}]\cup
[\tau_\mathrm{Strong},\infty[$ by the integral covering the whole
dominion $]-\infty,\infty[$, we commit a relative error of order
$2m/b \ll 1$. We can finally express the whole deflection vector
as
\begin{eqnarray}
\hat\bfalpha & = & \hat\bfalpha_\mathrm{BH} +
\hat\bfalpha_\mathrm{Q} \label{totalpha} \\
\hat\bfalpha_\mathrm{Q} & \simeq & 2\int\limits^{\infty}_{-\infty}
\nabla_\perp \varphi_\mathrm{Q} d\tau. \label{intalphaQ}
\end{eqnarray}
So we have recovered a superposition principle between the strong
deflection due to the black hole and the weak deflection due to
the external field. This has been possible since the perturbation
of the photon trajectory arising in the strong field regime is
very small compared to that arising in the weak field regime. As
we shall see in the following sections, the consequence of this
crucial fact is that the lens equation preserves a very simple
structure, allowing a straightforward analysis of the effects of
the external field on the relativistic images.

\section{Perturbed Schwarzschild lens}
\label{sec:standard}

\subsection{Standard configuration}

Since the strong deflection angle due to the black hole,
$\hat\bfalpha_\mathrm{BH}$, and the weak deflection angle due to
the external field, $\hat\bfalpha_\mathrm{Q}$, can be added, the
lens equation will take a very simple form. However, the lens
equation (\ref{cr1}) and its specifications in the standard
(\ref{cr3}) and retro-lensing (\ref{cr3re}) geometries can not be
directly generalized to the presence of an external field. The
superposition principle was derived assuming that total deflection
occurs by steps. After the photons wind around the black hole,
they quickly reach the region where the only weak gravitational
field is effective and eventually approach the asymptotic
trajectory to the observer with impact parameter $u$. In other
words, the impact parameter $u_\mathrm{BH}$ which enters in the
expression for the deflection angle due to the black hole in
Eq.~(\ref{cr4}) differs from the impact parameter $u$ as seen from
the observer after the action of the perturber.
Equation~(\ref{cr1}) was explicitly derived assuming
$u_\mathrm{BH} = D_\mathrm{L}\theta$, which is no longer true when
an external deflection is present, and can not be directly
applied.

Let us define a coordinate system centered on the black hole and
with the $z$-axis oriented towards the observer. Then let us
consider a photon coming out from the black hole with an impact
parameter $u_\mathrm{BH}$. We shall indicate its projection on the
lens plane by the vector $\bfu_\mathrm{BH}$. As the photon travels
towards the observer, it will still be deflected by the external
field. Its trajectory can be thus parameterized as
\begin{equation}
\label{stan1}
\mathbf{x}(\tau)=\mathbf{x}_\mathrm{O}+\mathbf{\dot
x}_\mathrm{O}(\tau-\tau_\mathrm{O})-2\int
\limits_\tau^{\tau_\mathrm{O}} d\tau'
\int\limits_{\tau'}^{\tau_\mathrm{O}} \nabla_\perp
\varphi_\mathrm{Q} d\tau'',
\end{equation}
where $\mathbf{x}_\mathrm{O}=(0,0,D_\mathrm{L})$ is the observer
position; $\mathbf{\dot x}_\mathrm{O}=(-\theta_1,-\theta_2,1)$ is
the tangent vector when the photon reaches the observer (to first
order in $\bftheta$) and $\tau_\mathrm{O}\simeq D_\mathrm{L}$ is
the value of the affine parameter in which the photon reaches the
observer. Note that here we are only including the weak field
shift due to the external field, as the black hole deflection is
nearly ineffective in the weak field region. We will come back
later to this point for its full clarification. Furthermore, we
are only interested in how the perturber affects the relation
between the asymptotic trajectory and the impact parameter at the
black hole position.

The integral in Eq.~(\ref{stan1}) can be calculated along the
unperturbed trajectory
$\mathbf{x}^{(0)}(\tau)=(u^\mathrm{BH}_1,u^\mathrm{BH}_2,\tau)$.
Moreover, the integrand is significant only for $\tau\lesssim b$
so the result does not change if we push the upper integration
limit $\tau_\mathrm{O}$ to infinity. So, evaluating
$\mathbf{x}(\tau)$ at $\tau=0$ and identifying $\bfu_\mathrm{BH}$
with $\mathbf{x}(0)$, we finally have
\begin{eqnarray}
\bfu_\mathrm{BH} & = &\mathbf{x}(0) =  D_\mathrm{L} \bftheta - \Delta
\mathbf{u}_\mathrm{O} , \label{LensO}
\\
\Delta \mathbf{u}_\mathrm{O}& \equiv & 2\int \limits_0^{\infty}
dz' \int\limits_{z'}^{\infty} \nabla_\perp
\varphi_\mathrm{Q}(u^\mathrm{BH}_1,u^\mathrm{BH}_2,z'') dz''.
\label{Delta uO}
\end{eqnarray}
Since $\mathbf{u} = D_\mathrm{L} \bftheta $, we see how the
presence of an external perturbation modifies the relation between
angles and impact parameters. The lens and the relativistic images
it produces are shifted by $\Delta \bfu_\mathrm{O}/D_\mathrm{L}$
in the observer sky w.r.t. their true position relative to the
black hole.

Let us estimate the order of magnitude of the shift term in
Eq.~(\ref{Delta uO}). By definition$, \Delta
\mathbf{u}_\mathrm{O}$ is the spatial shift due to the deflection
action of the perturber between the black hole and the observer.
As before, the transverse gradient introduces a factor $b^{-1}$,
while the two integrals multiply the result by $b^2$. We thus have
$|\Delta \mathbf{u}_\mathrm{O}| \approx \varphi_\mathrm{Q} b$.
This quantity might be comparable with $u_\mathrm{BH}$ itself. To understand
this, as perturbing body let us consider a second black hole with
mass $m_2$ at distance $b$. Then $|\Delta
\mathbf{u}_\mathrm{O}|\approx 2m_2$ and if the second black hole
has a mass similar to the first one, $u_\mathrm{BH} \approx 2 m_1 \approx 2
m_2 \approx |\Delta \mathbf{u}_\mathrm{O}|$.

The reader may wonder why in the standard weak field limit we do
not need to evaluate this kind of shift of the lens plane. Indeed,
there is a big difference. In fact, in the weak field, the scale
of variations in the deflection angle is what we have called $b$
and is very large if the photons pass very far from the black
hole. Then, a shift of order $\varphi_\mathrm{Q}b$ in the lens
plane is absolutely negligible on the scale of the deflection
$b$. It is practically a higher order term in a weak deflection
expansion. For relativistic images, where we need a resolution of
the order of the Schwarzschild radius $2m_1$, this shift of the
apparent lens position is of the same order of the relative
positions of the images.

Another point to clarify is the fact that we are assuming that the
whole deflection by the black hole takes place in a region of
order $2m_1$, neglecting any weak field deflection by the black
hole. Indeed, the weak field deflection induced by the black hole
is governed by the transverse gradient of its potential. Yet the
photon is coming out with a very small impact parameter, so the
transverse gradient is of order $2m_1 u_\mathrm{BH}/b^3 \approx
(2m_1)^2/b^3$ while the transverse gradient of the potential of
the perturber is $\varphi_\mathrm{Q}/b$, so that the first is of
higher order in the gravitational potential w.r.t. the second. So,
we can safely assume that the photon coming out from the black
hole reaches the asymptotic trajectory in few Schwarzschild radii
and suffers no more bending by the black hole. This can be
explicitly verified plotting the trajectory of a photon
experiencing a very large deflection.

To get the lens equation, we must relate $\bftheta$  to $\bfbeta$.
This can be done adding the information on the photon path between
the source and the black hole in analogy to Eq.~(\ref{LensO}). In
fact, the photon trajectory from the source to the black hole can
be parameterized by
\begin{equation}
\mathbf{x}(\tau)=\mathbf{x}_\mathrm{S}+\mathbf{\dot
x}_\mathrm{S}(\tau-\tau_\mathrm{S})-2\int
\limits^\tau_{\tau_\mathrm{S}} d\tau'
\int\limits^{\tau'}_{\tau_\mathrm{S}} \nabla_\perp
\varphi_\mathrm{Q} d\tau'', \label{xtauS}
\end{equation}
where
$\mathbf{x}_\mathrm{S}=(D_\mathrm{S}\beta_1,D_\mathrm{S}\beta_2,-D_\mathrm{LS})$
is the source position; $\mathbf{\dot
x}_\mathrm{S}=(\hat\alpha_1-\theta_1,\hat\alpha_2-\theta_2,1)$ is
the tangent vector when the photon leaves the source (to first
order in $\bftheta$ and $\hat\bfalpha$) and $\tau_\mathrm{S}\simeq
-D_\mathrm{LS}$ is the value of the affine parameter in which the
photon leaves the source. Note that the deflection vector entering
here is the total deflection given by Eq. (\ref{totalpha}).

Let us determine the point where the photon hits the lens plane.
Since we are neglecting the effects of the external field when the
photon winds around the black hole, the motion around the black
hole is still planar. Then, the incidence point coincides with the
exit point $\bfu_\mathrm{BH}$. We thus set
$\mathbf{x}(0)=\bfu_\mathrm{BH}$ in Eq.~(\ref{xtauS}), so that
\begin{eqnarray}
\bfu_\mathrm{BH} & = & \mathbf{x}(0) = D_\mathrm{S}\bfbeta+D_\mathrm{LS}(\hat\bfalpha-\bftheta)-
\Delta \mathbf{u}_\mathrm{S} \label{LensS}
\\
\Delta \mathbf{u}_\mathrm{S} & \equiv & 2\int \limits^0_{-\infty}
dz' \int\limits^{z'}_{-\infty} \nabla_\perp
\varphi_\mathrm{Q}(u^\mathrm{BH}_1,u^\mathrm{BH}_2,z'') dz''.
\label{Delta uS}
\end{eqnarray}
Now, eliminating $\bfu_\mathrm{BH}$ between Eqs.~(\ref{LensO}) and
(\ref{LensS}), we get the perturbed Schwarzschild lens equation in
the standard geometry
\begin{equation}
\bfbeta = \bftheta - \frac{D_\mathrm{LS}}{D_\mathrm{S}} \left[
\hat{\bfalpha}_\mathrm{BH} + \hat\bfalpha_\mathrm{Q}\right] +
\frac{\Delta \bfu_\mathrm{S}-\Delta
\bfu_\mathrm{O}}{D_\mathrm{S}}. \label{StandardtTheta}
\end{equation}
Apart from the presence of the perturber deflection angle
$\hat\bfalpha_\mathrm{Q}$, we have the two shifts $\Delta
\bfu_\mathrm{S}$ and $\Delta \bfu_\mathrm{O}$. By their
definition, it is easy to verify that
\begin{equation}
\Delta \bfu_\mathrm{S}-\Delta \bfu_\mathrm{O} =
D_\mathrm{LS}\hat{\bfalpha}_Q -2 \int
\limits_{-D_\mathrm{LS}}^{D_\mathrm{L}} d \tau'
\int \limits_{\tau'}^{D_\mathrm{L}} \nabla_\perp
\varphi_\mathrm{Q} d\tau''.
\end{equation}
The total shift is related to the discrepancy between the position
of the center of the perturber and the plane of the black hole. It
can be effective if the external field mostly intervenes before or
after the black hole and it is of higher order and can be
neglected in the lens equation only if the distance between black
hole and perturber along the line of sight is of the same order of
the Schwarzshild radius of the black hole. If the deflector
potential has the symmetry
$\varphi_\mathrm{Q}(u^\mathrm{BH}_1,u^\mathrm{BH}_2,z)=\varphi_\mathrm{Q}(u^\mathrm{BH}_1,u^\mathrm{BH}_2,-z)$,
then the shift is null.

The perturbed lens equation~(\ref{StandardtTheta}) reproduces the
vector generalization of Eq.~(\ref{cr3}) when the external field
vanishes. The black hole deflection vector
$\hat{\bfalpha}_\mathrm{BH}$ in standard lensing reduces to the
offset $\Delta \hat{\bfalpha}_n^\mathrm{BH}$, introduced in
Section~\ref{sec:lenseq}, in its obvious vector generalization.
Needless to say, both deflections $\Delta
\hat{\bfalpha}_{n}^\mathrm{BH}$ and $\hat\bfalpha_\mathrm{Q}$ must
be evaluated in $\bfu_\mathrm{BH}=\bfu-\Delta
\mathbf{u}_\mathrm{O}$.

At this point it is worth to eliminate $\bftheta$ throughout the
lens equation and use $\mathbf{u}_\mathrm{BH}$ instead. However, we must take
care of the spatial dependence of $\Delta
\mathbf{u}_\mathrm{O}$ and $\Delta \mathbf{u}_\mathrm{S}$. We can expand them around
$\mathbf{u}_\mathrm{BH}=0$ and write
\begin{eqnarray}
\Delta \mathbf{u}_\mathrm{O} & =& \Delta
\mathbf{u}_\mathrm{O}^{(0)}+\hat J_\mathrm{O} \mathbf{u}_\mathrm{BH}\\
\Delta \mathbf{u}_\mathrm{S} & =& \Delta
\mathbf{u}_\mathrm{S}^{(0)}+\hat J_\mathrm{S} \mathbf{u}_\mathrm{BH}
\end{eqnarray}
where $\hat J_\mathrm{O}$ and $\hat J_\mathrm{S}$ are the Jacobian
matrices of $\Delta \mathbf{u}_\mathrm{O} $ and $\Delta
\mathbf{u}_\mathrm{S}$ evaluated at $\mathbf{u}_\mathrm{BH}=0$.

In the same way, we can express the perturber deflection angle as
in the classical Chang \& Refsdal lens, expanding to first order
in $\bfu$
\begin{eqnarray}
\label{cr10} && \bfalpha_Q \equiv \frac{D_{LS}}{D_S}
\hat{\bfalpha}_Q
= \bfalpha_Q (0) +\hat J_\mathrm{Q} \mathbf{u}_\mathrm{BH} \\
&& \hat J_\mathrm{Q}= \frac{1}{D_\mathrm{L}}\left(
\begin{array}{cc}
\kappa_\mathrm{Q} + \gamma_\mathrm{Q} & 0 \\ 0 & \kappa_\mathrm{Q}
- \gamma_\mathrm{Q}
\end{array}
\right) ,
\end{eqnarray}
where $\kappa_\mathrm{Q}$ and $\gamma_\mathrm{Q}$ are the
convergence and the shear of the perturber at the location of the
black hole, respectively, and the orientation is chosen to
diagonalize the tidal matrix $\hat J_\mathrm{Q}$ \cite{sef}.

After all these considerations on the terms of Eq.
(\ref{StandardtTheta}), we can write the lens equation as
\begin{equation}
\bfbeta = D_\mathrm{L}^{-1} [ \mathbf{u}_\mathrm{BH} + \Delta
\mathbf{u}^{(0)}_\mathrm{O}+ \hat J_\mathrm{O}
\mathbf{u}_\mathrm{BH}] - \frac{D_\mathrm{LS}}{D_\mathrm{S}}\Delta
\hat{\bfalpha}_{n}^\mathrm{BH} (\mathbf{u}_\mathrm{BH}) -
\bfalpha_Q (0) - \hat J_\mathrm{Q} \mathbf{u}_\mathrm{BH}
+D_\mathrm{S}^{-1} [\Delta \mathbf{u}^{(0)}_\mathrm{S}+\hat
J_\mathrm{S} \mathbf{u}-\Delta \mathbf{u}^{(0)}_\mathrm{O}-\hat
J_\mathrm{O} \mathbf{u}] \label{Lens without shift}
\end{equation}
We can englobe all constant terms in $\bfbeta$, redefining the
origin of the source plane. Rather than introducing a new
notation, for simplicity we will continue to use $\bfbeta$ and
drop the constant quantity $\Delta \bfbeta$ from the right hand
side of Eq.~(\ref{Lens without shift}),
\begin{equation}
\Delta \bfbeta  \equiv
\frac{D_\mathrm{LS}}{D_\mathrm{L}D_\mathrm{S}} \Delta
\mathbf{u}^{(0)}_\mathrm{O} - \bfalpha_Q (0) \label{shift}
+\frac{1}{D_\mathrm{S}} \Delta\mathbf{u}^{(0)}_\mathrm{S}
  \simeq  - \bfalpha_Q (0).
\end{equation}
Note that the shift in the source plane is largely dominated by
$\bfalpha_Q (0)$, which is of order $\varphi_\mathrm{Q}$, while
all other shift terms are at most of order
$\varphi_\mathrm{Q}b/D_\mathrm{S}$ and we remind that we are
assuming the scale of variation of the external field $b$ to be
much smaller than all distances among source, lens and observer.

Similarly, all spatial variations of the shift $\Delta
\mathbf{u}_\mathrm{O}$  are negligible w.r.t. the deflection angle
spatial variation, so that they too can be neglected from the lens
equation, which reduces to
\begin{equation}
\bfbeta = D_\mathrm{L}^{-1} \mathbf{u}_\mathrm{BH} -
\frac{D_\mathrm{LS}}{D_\mathrm{S}}\Delta
\hat{\bfalpha}_{n}^\mathrm{BH}(\mathbf{u}_\mathrm{BH}) -\hat J_\mathrm{Q} \mathbf{u}_\mathrm{BH} .
\end{equation}
As anticipated before, the shifts play no role in the lens
equation. However, $\Delta \mathbf{u}_\mathrm{O}$ intervenes in
the relation between $\bftheta$ and $\bfu_\mathrm{BH}$ and must be
taken into account to describe the correct position of the images
in the sky.

$\Delta \hat{\bfalpha}_{n}^\mathrm{BH}$ can be expressed by Eq.
(\ref{cr6}) where now $\epsilon = (u_\mathrm{BH} -
D_\mathrm{L}\theta_n^0)/(D_\mathrm{L}\theta_n^0)$ and $\theta_n^0$
is still given by Eq. (\ref{cr5}). Writing the components of
$\mathbf{u}_\mathrm{BH}$ as
$D_\mathrm{L}\theta_{n}^0(1+\epsilon)(\cos \phi,\sin \phi)$ and
those of $\bfbeta$ as $(\beta_1,\beta_2)$, the lens equation can
be written to zero order in $\epsilon$ as
\begin{eqnarray}
\beta_1 & = & \theta^0_{n}\left[  1-\kappa_\mathrm{Q}
-\gamma_\mathrm{Q}-
\frac{\epsilon}{\epsilon_{n}^\mathrm{E}} \right]\cos \phi ,  \label{cr13}\\
\beta_2 & = & \theta^0_{n}\left[  1-\kappa_\mathrm{Q}
+\gamma_\mathrm{Q}- \frac{\epsilon}{\epsilon_{n}^\mathrm{E}}
\right]\sin \phi, \label{cr14}
\end{eqnarray}
where $\epsilon_{n}^\mathrm{E}$ is given by Eq. (\ref{cr8}).

The perturbed lens equation has an amazingly simple form, which
encourages its complete investigation. Let us study critical
curves and caustics. The determinant of the Jacobian matrix is
\begin{equation}
\label{cr15} J = \frac{1}{\left(
{\epsilon_{n}^\mathrm{E}}\right)^2}\left[ \epsilon - (1
-\kappa_\mathrm{Q} +\gamma_\mathrm{Q} \cos
2\phi)\epsilon_{n}^\mathrm{E}\right].
\end{equation}
Setting $J=0$, the critical curves are found as
\begin{equation}
\label{cr16} \epsilon = (1-\kappa_\mathrm{Q} +\gamma_\mathrm{Q}
\cos 2 \phi)\epsilon_{n}^\mathrm{E},
\end{equation}
which, in terms of $\bftheta$, correspond to
\begin{equation}
\bftheta = D_\mathrm{L}^{-1}\Delta \mathbf{u}_\mathrm{O}^{(0)}+
\theta_{n}^0[1+(1-\kappa_\mathrm{Q} +\gamma_\mathrm{Q} \cos 2
\phi)\epsilon_{n}^\mathrm{E}] (\hat I+\hat J_\mathrm{O}) \left(
\begin{array}{c}
\cos \phi \\ \sin \phi \end{array} \right),
\end{equation}
where we have denoted the two-dimensional identity matrix by $\hat
I$. The circular symmetry of the Einstein rings is broken both by
$\gamma_\mathrm{Q}$ and by $\hat J_\mathrm{O}$, which give
contributions of similar relevance here. The convergence
$\kappa_\mathrm{Q}$ only acts as an isotropic focusing factor.
Note that the perturbations to the critical curves stay small even
if $\kappa_\mathrm{Q}$ and $\gamma_\mathrm{Q}$ are larger than 1.
This is quite different from the weak field Chang \& Refsdal lens,
where the critical curves and the caustics change their topology
when the convergence and the shear become larger than 1.

The perturbed critical curves are mapped in symmetric
diamond-shaped caustics centered at $\bfbeta=0$, with parametric
equations
\begin{equation}
\left\{ \begin{array}{c} \beta_1 \\ \beta_2 \end{array} \right. =
-2 \gamma_\mathrm{Q} \ \theta^0_{n} {\times} \left\{
\begin{array}{c} \cos^3 \phi \\ \sin^3 \phi
\end{array}
\right. .
\end{equation}
The half-width of the caustics $\gamma_\mathrm{Q} \ \theta^0_{n}$
decreases with the number of loops, tending to the asymptotic
value $\gamma_\mathrm{Q} \ \theta_\mathrm{m}$. In principle, a
point-like source might be inside the first caustics but outside
the caustics corresponding to large values of $n$. The shift of
the caustics from the optical axis is given by Eq.~(\ref{shift})
and is the same for all of them (it even coincides at the first order with the shift
of the weak field caustic) and amounts to the zero-order term in
the expansion of the deflection angle due to the external field
$\bfalpha_Q (0)$. This is very different from the effect of an
intrinsic angular momentum of the black hole
\cite{boz03,boz+al05}, where the size and the shift of the
caustics increases linearly with $n$. So, the degeneracy between
external shear and intrinsic spin of the lens can be
broken, in principle, observing relativistic images of different
order in $n$.

The solutions of the general lens equation can be found by
reducing it to a one-dimensional form. Adding the squares of $\cos
\phi$ and $\sin \phi$ yields a fourth-order equation for
$\epsilon$ that can be solved by standard methods. For each number
of loops $n$ around the black hole, the maximum number of images
is four. The expressions of the solutions in the general case are
lengthy and we do not report them here. Basic properties of image
multiplicity can be obtained by considering images for a source on
one of the axes, for example $\beta_2=0$. We have the following
possibilities. If $\sin \phi =0$, then $\beta_1 =\pm \theta_{n}^0
(1 - k_\mathrm{Q} - \gamma_\mathrm{Q} -
\epsilon/\epsilon_{n}^\mathrm{E})$. Two images form at $\phi=0$
and $\phi=\pi$ , with
\begin{equation}
\label{cr17} \epsilon_{0,\pi} =   \left[
(1-k_\mathrm{Q}-\gamma_\mathrm{Q}) \mp
\frac{\beta_1}{\theta_{n}^0} \right] \epsilon_{n}^\mathrm{E} .
\end{equation}
This couple of series of images always exists and matches the
unperturbed solutions when the external perturbation is switched
off. If $\sin \phi \neq 0$, then it must be $ (1 - k_\mathrm{Q}
+\gamma_\mathrm{Q} -\epsilon/\epsilon^E_{n}) = 0$. For a source
inside a caustic, when the condition $\beta_1 < 2
\gamma_\mathrm{Q} \theta_{n}^0$ is fulfilled, an additional couple
of images is produced, whereas sources outside it have only two.
The new sequences of images form at
\begin{eqnarray}
\label{cr18}
\epsilon & = & ( 1 - k_\mathrm{Q} +
\gamma_\mathrm{Q})\epsilon^\mathrm{E}_{n}, \\
 \phi & = & \pm \arccos
\left(-\frac{\beta_1}{2\gamma_\mathrm{Q} \theta^0_{n}}\right).
\end{eqnarray}

Note that the caustic is fully inverted w.r.t. the critical curve.
This means that if the source enters from the top-right side, the
new images form on the left-bottom side. This is different from
what happens in Kerr lensing \cite{boz+al05}, where the caustic is
only inverted on the left-right direction.

As usual, to translate these values of $\epsilon$ and $\phi$ into
positions in the observer sky, we must use Eq. (\ref{LensO}),
which gives
\begin{equation}
\bftheta = D_\mathrm{L}^{-1}\Delta \mathbf{u}_\mathrm{O}^{(0)}+
\theta_{n}^0(1+\epsilon) (\hat I+\hat J_\mathrm{O}) \left(
\begin{array}{c}
\cos \phi \\ \sin \phi \end{array} \right) .  \label{thetau}
\end{equation}

We can finally note that the total magnification of an image at
position $(\epsilon,\phi)$ is given by the inverse of the Jacobian
(\ref{cr15}), since the transformation from $\bfu_\mathrm{BH}$ to
$\bftheta$ through Eq. (\ref{LensO}) introduces negligible
corrections to the Jacobian (of order $\varphi_\mathrm{Q}$). It is
interesting to note that the two eigenvectors of the Jacobian
matrix keep a quasi-radial and a quasi-tangential direction. We
can thus distinguish a radial and a tangential magnification,
which respectively read
\begin{eqnarray}
&& \mu_\mathrm{r}= -\epsilon_{n}^\mathrm{E}\\
&& \mu_\mathrm{t}= -\frac{\epsilon_{n}^\mathrm{E}}{ \epsilon - (1
-\kappa_\mathrm{Q} +\gamma_\mathrm{Q} \cos
2\phi)\epsilon_{n}^\mathrm{E}}.
\end{eqnarray}
These expressions nicely reproduce those given in Refs.
\cite{oha89,boz+al01,boz02}, in the limit
$\kappa_\mathrm{Q},\gamma_\mathrm{Q} \rightarrow 0$, once we
susbstitute the expression of $\epsilon_{n}^\mathrm{E}$
(\ref{cr8}) and the definition of $\epsilon$. Note that the radial
magnification is always positive ($\epsilon_{n}^\mathrm{E}<0$ by
its definition), while the tangential magnification is positive if
the image is outside the critical curve and negative otherwise. It
diverges for degenerate images on the critical curve. Thus, as in
all other cases in which relativistic images are highly magnified,
they appear as elongated and very thin tangential arcs
\cite{boz+al05}.

\subsection{Retro-lensing configuration}

The generalization of the retro-lensing equation (\ref{cr3re}) to
the presence of an external field proceeds in the same way.
Equation~(\ref{LensO}) is still valid, while Eq. (\ref{LensS})
must be revised because the source is on the same side of the
lens. Starting from Eq. (\ref{xtauS}), we now have
$\mathbf{x}_\mathrm{S}=(D_\mathrm{S}\beta_1,D_\mathrm{S}\beta_2,D_\mathrm{LS})$,
$\mathbf{\dot x}_\mathrm{S}=
(\hat\alpha_1-\theta_1,\hat\alpha_2-\theta_2,-1)$ and
$\tau_\mathrm{S}\simeq -D_\mathrm{LS}$. Note that the
$z$-component of the deflection vector $\hat\alpha_3\simeq-2$ is
responsible of the sign flip in the $z$-component of the tangent
vector.

Moreover, the photon must hit the lens plane at the point
$-\bfu_\mathrm{BH}$ in order to come back from the point
$\bfu_\mathrm{BH}$. Then we need to integrate along the
unperturbed trajectory
$\mathbf{x}^{(0)}(\tau)=(-u^\mathrm{BH}_1,-u^\mathrm{BH}_2,-\tau)$.
Evaluating Eq. (\ref{xtauS}) at $\tau=0$, we have
\begin{eqnarray}
-\mathbf{u}_\mathrm{BH} & = &
\mathbf{x}(0)=D_\mathrm{S}\bfbeta+D_\mathrm{LS}(\hat\bfalpha_\perp-\bftheta)-
\Delta \mathbf{u}_\mathrm{S} \label{LensSre}
\\
\Delta \mathbf{u}_\mathrm{S} & \equiv & 2\int \limits_0^{\infty}
dz' \int\limits_{z'}^{\infty} \nabla_\perp
\varphi_\mathrm{Q}(-u^\mathrm{BH}_1,-u^\mathrm{BH}_2,z'') dz'',
\label{Delta uSre}
\end{eqnarray}
where we have indicated by $\hat\bfalpha_\perp$ the projection of
the deflection vector on a plane perpendicular to the $z$-axis.
Compare these relations with Eqs. (\ref{LensS}) and (\ref{Delta
uS}).

Combining Eq. (\ref{LensSre}) with Eq. (\ref{LensO}), we get the
perturbed Schwarzschild retro-lens equation
\begin{equation}
\bfbeta = -\left( 1+2 \frac{D_\mathrm{LS}}{D_\mathrm{S}} \right)
\bftheta - \frac{D_\mathrm{LS}}{D_\mathrm{S}}
 \left[ \hat{\bfalpha}_\perp^\mathrm{BH} +
 \hat\bfalpha_\mathrm{Q}\right]+ \frac{\Delta \bfu_\mathrm{S}+\Delta
 \bfu_\mathrm{O}}{D_\mathrm{S}}, \label{Retrotheta}
\end{equation}
where now $\hat{\bfalpha}_\perp^\mathrm{BH}=
-\Delta\hat{\bfalpha}_n^\mathrm{BH}$. The minus sign is easily
understood in the following way: the projected deflection vector
increases its modulus as $0<\hat{\alpha}_\mathrm{BH}<\pi/2$. Then,
it has modulus equal to unity at $\hat{\alpha}_\mathrm{BH}=\pi/2$
and then starts to decrease as
$\pi/2<\hat{\alpha}_\mathrm{BH}<\pi$. At
$\hat{\alpha}_\mathrm{BH}=\pi$,
$\hat{\bfalpha}_\perp^\mathrm{BH}=0$; with a positive
$\Delta\hat{\bfalpha}_n^\mathrm{BH}$ it starts to increase its
modulus again but its new direction is opposite to the original
one. After this consideration, we can appreciate how the perturbed
lens equation (\ref{Retrotheta}) reduces to the unperturbed
retro-lens equation (\ref{cr3re}) when the perturbation vanishes.

Now let us expand the shifts $\Delta \bfu_\mathrm{S}$ and $\Delta
\bfu_\mathrm{O}$ for small values of $u$. Comparing Eq.
(\ref{Delta uSre}) with Eq. (\ref{Delta uO}), we get
\begin{eqnarray}
\Delta \mathbf{u}_\mathrm{O} & =& \Delta
\mathbf{u}^{(0)}_\mathrm{O}+\hat J_\mathrm{O} \mathbf{u}_\mathrm{BH}\\
\Delta \mathbf{u}_\mathrm{S}& = & \Delta
\mathbf{u}^{(0)}_\mathrm{O}-\hat J_\mathrm{O}
\mathbf{u}_\mathrm{BH}.
\end{eqnarray}
In fact, the two shifts are calculated along very similar paths.
Thus, their first order expansion in $\bfu_\mathrm{BH}$ has the
same zero order term and opposite first order terms.

Something similar happens when we calculate the deflection angle
of the perturber $\hat\bfalpha_\mathrm{Q}$, since the incident and
exit paths are very close each other
\begin{equation}
\hat\bfalpha_\mathrm{Q}= 4\int\limits_{0}^{\infty} \nabla_\perp
\varphi_\mathrm{Q}(0,0,z) dz+ 2\int\limits_{0}^{\infty}
\frac{\partial \nabla_\perp \varphi_\mathrm{Q}}{\partial
\bfu_\mathrm{BH}} \bfu_\mathrm{BH}~ dz- 2\int\limits_{0}^{\infty}
\frac{\partial \nabla_\perp \varphi_\mathrm{Q}}{\partial
\bfu_\mathrm{BH}} \bfu_\mathrm{BH}~ dz, \label{alphaQre}
\end{equation}
where the symbol $\partial \nabla_\perp
\varphi_\mathrm{Q}/\partial \bfu_\mathrm{BH}$ indicates the
Jacobian of the transverse gradient of the potential w.r.t. to
$\bfu_\mathrm{BH}$.

While the zero-order contributions in the approach to the black
hole and the way back sum up, the first order contributions in
$\bfu_\mathrm{BH}$ cancel out. We thus conclude that, up to
octopole terms, the perturbed Schwarzschild lens in the
retro-lensing geometry is equivalent to a shifted Schwarzschild
lens. Alas, the octopole contributions are of the same order of
the contribution coming from the strong field region. Their
complete investigation requires a careful evaluation of the
perturbation of the photon trajectory in the strong field regime,
unnecessary in standard lensing. This higher order investigation
is beyond the scope of this work. So, we can conclude this section
saying that, up to perturbations of the order
$\varphi_\mathrm{Q}(2m)/b$, retro-lensing caustics are shifted but
remain point-like. The shift is generally different from that
suffered by standard lensing caustics and it is approximately
given by the first term on the right hand side of
Eq.~(\ref{alphaQre}). As regards the images and the critical
curves, everything proceeds as in the unperturbed case. The only
difference is that $\bftheta$ is given as a function of $\epsilon$
and $\phi$ by Eq.~(\ref{thetau}).

\section{Conclusions}
\label{sec:conc}

Super-massive black holes are supposed to be hosted by galactic
bulges, where the gravitational field of the dense stellar
environment may perturb gravitational lensing phenomenology. While
the perturbation of external fields on the Schwarzschild lens in
the weak deflection limit is well-known, our paper has clarified
the effects of a tidal perturbation on relativistic images
generated by photons winding one or more times around a
Schwarzschild black hole.

We have firstly given a form of the unperturbed lens equation more
general than that given in Ref. \cite{vi+el00}.

By a dimensional argument, we have then demonstrated that the
perturbation to the photon trajectory mainly arises in a weak
field regime, since the perturbation in the strong regime is of
higher order in the local curvature radius produced by the
perturbing body. On the other hand, the black hole deflection in
the weak field regime can be safely neglected, since the photon
trajectory is almost face-on and the transverse gradient of the
black hole potential is very small.

In this way, we have practically decoupled the two deflections.
This justifies the construction of the lens equation as a
three-steps process. First we have the weak deflection by the
perturber before the photon reaches the black hole, second we have
the strong deflection by the black hole and last we have another
weak deflection by the perturber. We have explicitly built the
lens equation for the two most interesting cases: standard and
retro-lensing.

For the standard case (source behind the black hole) the perturbed
lens equation reveals that the infinite sequence of relativistic
Einstein rings near the photon sphere is slightly distorted and
the point-like caustics get a finite size, being shifted by a
constant amount. The size of the caustics decreases with the
number of loops, being of order of the size of the shadow of the
black hole multiplied by the external shear. A source inside a
diamond-shaped caustic generates two additional relativistic
images, as usual.

In the retro-lensing geometry, the perturbation of the external
field cancels out at the lowest order. In order to evaluate the
effect, it is necessary to go to the next order in $1/b$, which also
involves the perturbation of the photon trajectory in the strong
field region around the central black hole.

The present work certainly does not exhaust all possible cases of tidal
distortion, since we have restricted to weak perturbers in a
static situation. However, the most physically relevant cases are
clearly included in this class. In fact, the gravitational field
at the Galactic center generated by the environment surrounding
the supermassive black hole should be reasonably described in the
weak field approximation. Gravitational lensing by Sgr~A* could be
a very accurate tool to track the existence of a non-vanishing
shear in the field generated by the environment. In other
galaxies, where sometimes the bulge hosts more than one
supermassive black hole, gravitational lensing may help to detect
the mass ratios of different black holes. In any case, given the
ubiquity of shear fields, the study presented here fills an
important gap in the literature about gravitational lensing in
strong gravitational fields.

\begin{acknowledgments}
We thank all the participants of the workshop `Gravitational Lensing
in the Kerr Spacetime Geometry', held at the American Institute of
Mathematics in Palo Alto (CA), for invaluable discussions and precious
interaction, and in particular S.~Frittelli and A.O.~Petters
for the kind invitation.

V.B. is supported by MIUR through PRIN 2004 `Astroparticle Physics'
and by research fund of the Salerno University. M.S. is supported by
the Swiss National Science Foundation and by the Tomalla Foundation.

\end{acknowledgments}

\bibliography{cr}

\end{document}